\documentclass[sigconf]{acmart}

\usepackage{booktabs} 

\makeatletter                   
\def\mdseries@tt{m}             
\makeatother                    
\usepackage[plain]{fancyref}
\usepackage[draft=true]{minted} 
\usepackage{color}
\usepackage{hyperref}           
\hypersetup{
    colorlinks=true,
    linkcolor=blue,
    filecolor=red,      
    urlcolor=magenta,
    breaklinks=true,            
}
\usepackage{breakurl}           

\setcopyright{none}

\acmConference[CHI PLAY 2020]{CHI PLAY 2020}{November 2020}{Ottawa, Canada}
\copyrightyear{2020}

\setcopyright{none}
\renewcommand\footnotetextcopyrightpermission[1]{}
\begin{document}
\sloppy                         

\title[Understanding Learners' Problem Solving Strategies in Concurrent and Parallel Programming]{Understanding Learners' Problem-Solving Strategies in Concurrent and Parallel Programming: A Game-Based Approach}

\author{Jichen Zhu}
\email{jichen@drexel.com}
\affiliation{%
  \institution{Drexel University}
  \city{Philadelphia}
  \country{USA}
}

\author{Katelyn Alderfer}
\email{kmb562@drexel.edu}
\affiliation{%
  \institution{Drexel University}
  \city{Philadelphia}
  \country{USA}
}

\author{Brian Smith}
\email{bks59@drexel.edu}
\affiliation{%
  \institution{Drexel University}
  \city{Philadelphia}
  \country{USA}
}

\author{Bruce Char}
\email{charbw@drexel.edu}
\affiliation{%
  \institution{Drexel University}
  \city{Philadelphia}
  \country{USA}
}

\author{Santiago Onta\~{n}\'{o}n}\authornote{Currently at Google.}
\email{so367@drexel.edu}
\affiliation{%
  \institution{Drexel University}
  \city{Philadelphia}
  \country{USA}
}

\renewcommand{\shortauthors}{Zhu, et al.}

\begin{abstract}
Concurrent and parallel programming (CPP) is an increasingly important subject in Computer Science Education. However, the conceptual shift from sequential programming is notoriously difficult to make. Currently, relatively little research exists on how people learn CPP core concepts. This paper presents our results of using {\em Parallel}, an educational game about CPP, focusing on the learners' self-efficacy and how they learn CPP concepts. Based on a study of 44 undergraduate students, our research shows that (a) self-efficacy increased significantly after playing the game; (b) the problem-solving strategies employed by students playing the game can be classified in three main types: trial and error, single-thread, and multi-threaded strategies, and (c) that self-efficacy is correlated with the percentage of time students spend in multithreaded problem-solving.

\end{abstract}

%
%


\maketitle


\section{Introduction}\label{sec:introduction}
Prior to 2006, most personal computers consisted of a single central processing unit (CPU) that was responsible for executing software. Today, most computing devices have multiple cores, requiring programmers to write software that explicitly takes advantage of the parallelism inherent across multiple cores. As a result, the concepts and skills associated with parallel computing are becoming a critical part of computer science education and computer science research. As important as this area is, even seasoned computer programmers have difficulty in the shift from sequential to concurrent and parallel programming. Thus it is important that researchers understand this conceptual shift and the learning needed to take place. 

Games have become an accepted media for education and training~\cite{gee2003video,prensky2003digital,randel1992effectiveness,shaffer2006computer}. Growing evidence shows that well-designed educational games not only sustain students' motivation for learning how to program, but also enhance the learning outcome~\cite{barnes2008game2learn,cliburn2006effectiveness,papastergiou2009digital,randel1992effectiveness,ibrahim2010students}. 

However, the majority of existing educational programming games target sequential programming and particularly for novice programmers~\cite{harteveld2014design, pirovano2014fuzzy}. Currently, not enough work has been conducted on how to students learn concurrent and parallel programming (CPP) concepts and how to scaffold learning in game-based learning environments. 

\begin{figure*}[tb]
    \begin{minipage}{\textwidth}
    \centering
    \raisebox{-0.5\height}{\includegraphics[width=0.81\textwidth]{level06.png}}
    \hspace*{0.1cm}
    \raisebox{-0.5\height}{\includegraphics[width=0.17\textwidth]{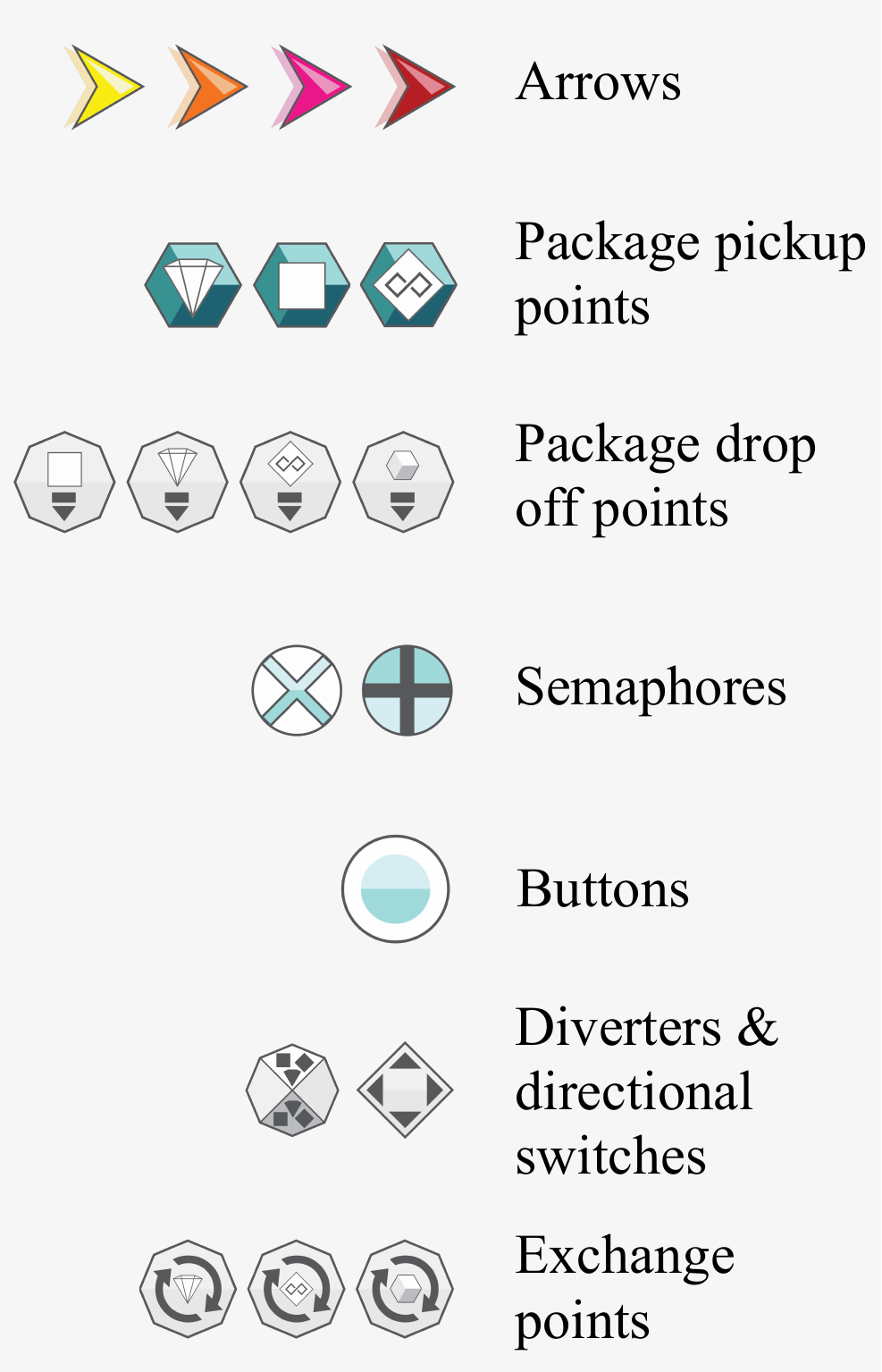}}
    \end{minipage}
    \caption{One of the {\em Parallel} levels used in our study, level 6.}
    \label{fig:level06}
\end{figure*}

Built on prior research on game-based learning, in this paper, we report our work on using {\em Parallel}\footnote{\url{https://github.com/santiontanon/Parallel}.} (Figure \ref{fig:level06}), an educational game about CPP, to facilitate learning as well as gathering empirical data on how students learn. In this paper, we attempt to understand the potential impact of our game on students' self-efficacy (one's belief that she can learn and accomplish new things).

This exploratory study aims to, through an educational game, further understanding of how students with sequential programming background learn concurrent and parallel programming. We do so by identifying the different problem-solving strategies they deploy when playing the game, and their relation with self-efficacy: how is self-efficacy affected by playing the game, and how does self-efficacy affect the way students play the game.  


In particular, this paper aims to answer two general questions:
\begin{itemize}
    \item How does {\em Parallel} impact the learners' level of self-efficacy, if at all?
    \item What kind of problem-solving strategies do the learners use in the context of playing {\em Parallel}?
\end{itemize}

In order to answer these questions, we conducted a user study at a university in a major U.S. city. We recruited CS undergraduate students ($n=44$) enrolled in an Operating System Course to play the {\em Parallel} game. Among them, we further observed a random sample of six students ($n=6$) using think-aloud to probe their problem-solving process while they played {\em Parallel}. Our results indicate that: (a) self-efficacy increased significantly after playing the game; (b) the problem-solving strategies employed by students can be classified in three main types ({\em trial and error}, {\em single-threaded problem solving}, and {\em multithreaded problem solving}); and (c) that self-efficacy is correlated with the percentage of time students spend in multithreaded problem solving, our target strategy.


Prior work in {\em Parallel}, has demonstrated how the game design allowed students to draw connections between game concepts and CPP concepts~\cite{zhu2019programming} and how we used AI techniques to model users and procedurally generate new levels \cite{valls2017graph, kantharaju2018tracing}. Detail details of the game can be found in \cite{zhu2019programming,ontanon2017designing}. Compared to our prior work on {\em Parallel}, this paper reports results from a new user study\footnote{The user study data reported in this paper is publicly available at: \url{https://github.com/santiontanon/parallel-study-data} }. The core contributions of this paper are on identifying how the game affects students' self-efficacy and identifying the different problem-solving strategies students use to solve CPP problems within the game. This paper is among the first studies to analyze individual undergraduate CS students' problem-solving actions directly. To the best of our knowledge, this is currently the only work that examines how students learn CPP concepts in the context of game-based learning.  

The remainder of this paper is structured as follows. Section~\ref{sec:background} presents related work on educational games on CPP, how students learn CPP, and self-efficacy. Section~\ref{sec:design} presents {\em Parallel}, the game used in our study. After that, Section~\ref{sec:methods} presents the design of the research study, and Section~\ref{sec:results} reports on the study results. The paper closes with discussion, conclusions, and directions for future work.

\section{Related Work}\label{sec:background}
This section describes related work in educational games for Concurrent and Parallel Programming (CPP), learning science literature on how students learn CPP and self-efficacy. 

\subsection{Existing Educational games for CPP}
A significant amount of literature exists on game-based learning for systematic thinking and programming skills. 
However, the majority of these educational programming games target sequential programming and particularly the introductory level~\cite{harteveld2014design, pirovano2014fuzzy,barnes2008game2learn,cliburn2006effectiveness,papastergiou2009digital,randel1992effectiveness,ibrahim2010students}. 
Since CPP requires problem-solving in a fundamentally different way from sequential programming, new designs and environments are needed to scaffold learning. 

Few CPP-related programming games exist compared to games for sequential programming. Exceptions include {\em Parallel Bots}, {\em Parallel Blobs}, {\em SpaceChem}, {\em Parapple} and {\em Deadlock Empire}. Most of these games, however, do not fully capture all the key parallel programming concepts. For example, {\em Parallel Bots} and {\em Parallel Blobs}~\cite{inggs2017learning}, which extend LightBot by adding multiple threads,
are deterministic and therefore do not cover the core CPP concept of non-determinism. {\em SpaceChem} does not contain non-determinism and only supports two threads, greatly reducing the complexity of the problem space. Finally, while {\em Parapple} is non-deterministic, it does not cover basic synchronization concepts such as semaphores or critical sections. Due to the missing core concepts and the limited complexity, the above-mentioned games at their current state cannot adequately represent real challenges in CPP and thus cannot train players to gain the necessary problem-solving skills. Our game, {\em Parallel}, is amongst the first attempts to fill in this gap.  

Among existing games on CPP, the one that covers all key core concepts as {\em Parallel} is {\em Deadlock Empire}. In this game, rather than programming, the player plays the role of the ``scheduler'', trying to find an execution order of two threads, which causes execution issues. 
Compared to {\em Parallel}, however, {\em Deadlock Empire} only supports two to three threads, and does not contain a visual representation of CPP concepts but rather asks the user to find schedules in actual programs, but wrapped in a background storyline.

\subsection{How Students Learn CPP Concepts}
It is well-known that students face significant challenges in learning CPP concepts. However, compared to the learning science literature on how students learn basic programming skills, relatively little is known about how they learn CPP. 

A significant amount of existing work focuses on how to improve the {\em teaching} of CPP concepts~\cite{hughes2005towards,ben1999thinking}. Research has been reported on what happened with various languages \cite{cartwright2010drhabanero}, custom-designed thread visualization tools \cite{carr2003threadmentor,lonnberg2011evaluating}, model checkers \cite{sadowski2011practical}, techniques using formal descriptions of concurrent behavior as program invariants \cite{carro2013model, gjessing2012teaching}, and pattern oriented software architectures \cite{schmidt2013producing}. While these are very useful tools, they can be more effective if they are informed by fundamental understandings of how students learn CPP concepts. 

Currently significantly less is known about the {\em learning} of CPP concepts. In the limited existing work, researchers have started to shed light on the important questions of how students learn CPP, what challenges they face and why.  Resnick\cite{resnick1996beyond} reported that students often approach CPP problems with an incorrect ``centralized mindset'', assuming that there is a leader or controller process coordinating the multiple threads. Observing elementary-school children, he identified three main types of programming bugs: problem-decomposition bugs, synchronization bugs, and object-oriented bugs \cite{resnick1990multilogo}. 

Kolikant observed that two approaches based among high school students. Students with no programming experience approached CPP problems from {\em a user's perspective}, focusing only what is directly perceivable. In comparison, students with more CS training tended to develop {\em a programmer's perspective}, allowing them to reason about the underlying synchronization mechanism at play \cite{kolikant2004learning}. 
Choi and Lewis \cite{choi2000study} analyzed errors from 180 programs in a senior-level undergraduate operating system course, finding that 30\% of them contained synchronization mistakes. This body of work highlight that a key learning challenge is synchronization, which is the focus of our game {\em Parallel}.

L{\"o}nnberg et al. has produced a substantial body of work to advance the understanding of how students learn CPP. Through interviewing students about how they develop and test concurrent programs, the researchers found that the common reasons why students encounter difficulties are over-reliance on trial-and-error and cursory testing without accounting for non-determinism~\cite{lonnberg2009students}. In a separate study focusing on the impact of a visualization tool to aid learning~\cite{lonnberg2011evaluating}, they identified the workflow of how students attempt to solve a concurrent program. They also found that the visualization tool was not sufficiently used by the students. They hypothesized that it was because the visualization tool was not directly tied to the debugging strategy. Built on their findings, we designed {\em Parallel} in a way that the students directly program and debug their synchronization mechanism on the visual levels itself. 

This paper aims to {\bf extend the literature} of how students learn CPP concepts by being among the first studies to analyze individual undergraduate CS students' problem-solving actions directly. This extends the literature that is primarily based on post-hoc students interviews after they completed assignments (e.g.,\cite{lonnberg2007students} and that is based on primarily discourse analysis of students conversations among each other while they work on solving a CPP program in a team setting (e.g., \cite{kolikant2004learning,lonnberg2011evaluating}). Among researches that examine how students program, work tend to focus on the learning outcome - the completed computer codes~\cite{choi2000study}. To the best of our knowledge, the only work that studied the problem-solving process at the level of how students actually program was Resnick's 1990 study on elementary-school children~\cite{resnick1990multilogo}. We extend this work by focusing on the undergraduate CS students population. We argue that our population has a strong need for learning CPP but we currently do not sufficiently understand their learning process. Finally, this is currently the only work that examines CPP learning in the context of game-based learning.



\subsection{Self-efficacy}

Self-efficacy can be defined as the ``belief in one's capabilities to organize and execute the courses of action required to produce given attainments''~\cite[p.~3]{bandura1977self}\cite{lent1996social}. In other words, self-efficacy can be equated to one's confidence in themselves or their ability to do something. Since Bandura's initial definition, studies have shown that self-efficacy has had a great deal of impact on and can be a predictor of a number of traits related to learning, including academic performance, motivation, and learning processes~\cite{ramalingam2004self, zimmerman2000self}. For example, Collins'~\cite{collins1982self} research on self-efficacy in math classrooms showed that despite student ability level, students with high self-efficacy completed more math problems correctly than those who had lower self-efficacy. Zimmerman et al. ~\cite{zimmerman1992self} used path analysis to show that academic self-efficacy directly influenced student achievement.

Work also exists on the connection and application of self-efficacy to educational games. Studies have shown that digital educational games can increase students' intrinsic motivation, student achievement, motivation, and self-efficacy~\cite{ketelhut2007impact,meluso2012enhancing,wang2010effects,jui2011game}. Programming is also a domain where self-efficacy appears to be connected to student learning. Several studies of self-efficacy and programming show that high student self-efficacy and comfort level serve as good predictors of how students will perform academically in their programming courses~\cite{ramalingam2004self,wilson2001contributing}. 

We chose to focus on self-efficacy, because (1) prior research has shown that self-efficacy can help with student motivation and their learning processes~\cite{bandura1977self,collins1982self,schunk2012motivation}; (2) it has been shown that there is a connection between self-efficacy and learning programming~\cite{bergin2005influence}; (3) self-efficacy can be a predictor for academic performance~\cite{ramalingam2004self, zimmerman2000self}; and finally (4) unlike {\em knowledge gain}, it can be directly measured via questionnaires. Moreover, although educational games have been observed to increase learners' self-efficacy~\cite{ketelhut2007impact,meluso2012enhancing,ramalingam2004self}, less is known in the context of CPP.

\section{Parallel}\label{sec:design}

{\em Parallel}~(see Figure \ref{fig:level06}) is a single-player 2D puzzle game designed to teach concurrent and parallel programming (CPP) core concepts. In {\em Parallel}, a player places semaphores and buttons in order to direct arrows to pick up packages to the designated delivery points. In essence, the player designs a {\em synchronization mechanism} to coordinate multiple threads executing at the same time. The target audience of {\em Parallel} is CS students who are interested in basic concepts in CPP. We designed {\em Parallel} to be used as supplementary material in a regular course curriculum or as an informal learning game. 

{\em Parallel} currently has 18 hand-authored levels with increasing difficulty. 
A more detailed description of the game and its design rationale can be found in Zhu et al. \cite{zhu2019programming} and Valls et al. \cite{valls2017graph} for its procedural content generation (PCG) component. Note that the PCG feature was not used in the study reported in this paper. In the rest of this section, we will only provide enough information about the game to contextualize the novel contribution of this paper.

\begin{figure*}[tb]
    \centering
    \includegraphics[width=1.0\textwidth]{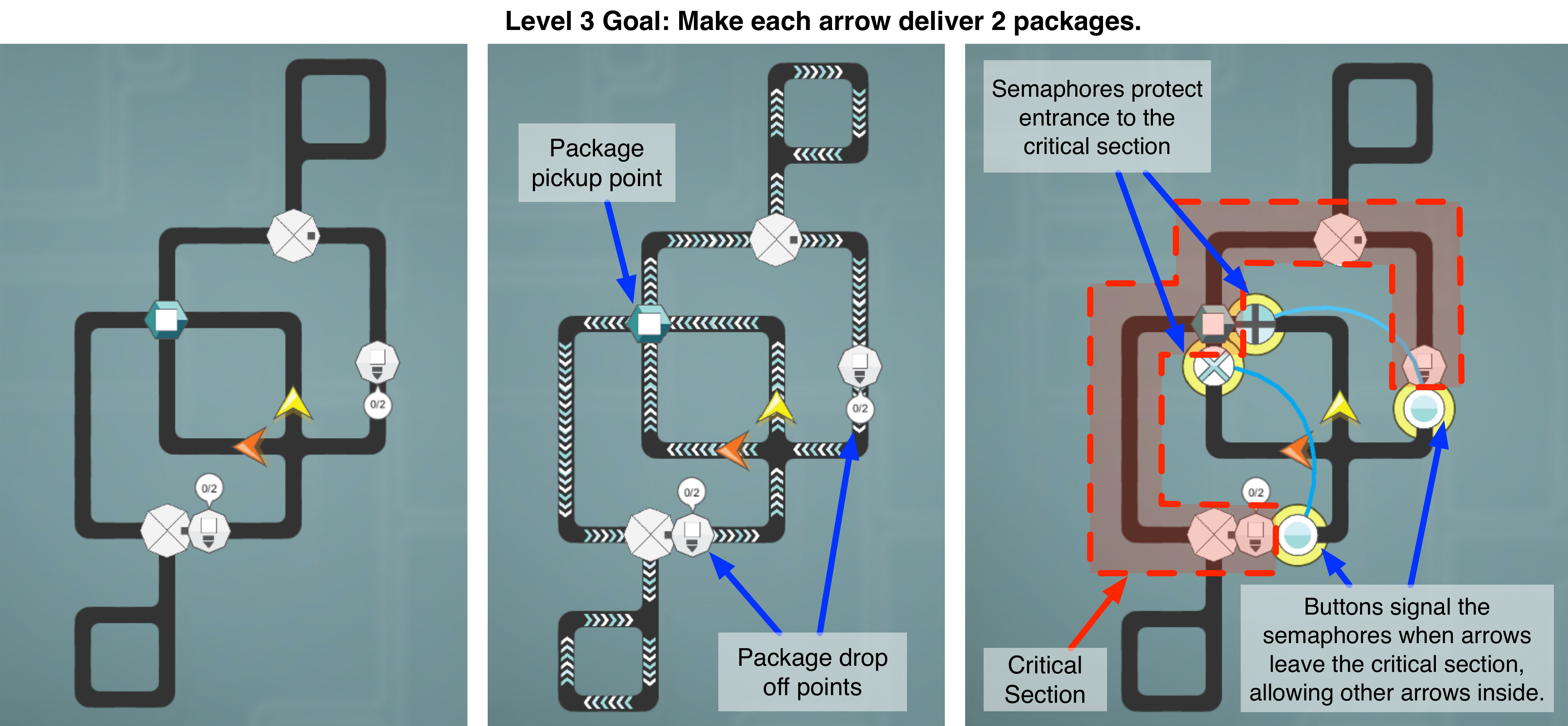}
\caption{Left: Screenshot of Level 3 about a level with two arrows (two threads), where students need to identify the critical section and block it. Middle: direction in which arrows move through the tracks. Right: solution to Level 3, highlighting the critical section, and how two semaphores and two buttons were used to block it.}
    \label{fig:level03}
\end{figure*}



\subsection{Connect CPP Concepts to Game Elements}

The game is more focused on concurrency topics (especially {\em non-determinism} and {\em synchronization}). It also includes parallelism topics such as {\em efficiency}. Table \ref{tbl:conceptMapping} summarizes the direct mapping between gameplay elements in {\em Parallel} and their counterpart CPP concepts.

\begin{table}[tb]\centering 
\caption{Mapping between game elements and CPP concepts.}
\label{tbl:conceptMapping} 
\begin{tabular}{|p{3cm}|p{4.8cm}|} \hline
{\bf Game Element} & {\bf Equivalent CPP Concept} \\ \hline \hline
Configuration of tracks & Program \\ \hline
Arrows & Threads of execution, a.k.a., threads \\ \hline
Packages & Resources \\ \hline
Semaphores & The ``wait'' operation of Semaphores \\ \hline
Buttons & The ``Signal'' operation of Semaphores \\ \hline
Diverters \& directional switches & Conditional statements \\ \hline
Exchange points & Message passing \\ \hline
\end{tabular}                            
\end{table}

{\em Arrows represent threads.} Arrows travel on the tracks and can be controlled by the player through semaphores and buttons. An arrow can carry any number of packages. 
The arrows exhibit {\em non-deterministic behavior} by traveling at randomized speeds, varying at randomized intervals. 

{\em Buttons and Semaphores represent the ``signal'' and ``wait'' operation of semaphores}. A button is triggered when an arrow passes through it. When the player links a button to a semaphore, and an arrow triggers the button, it will signal the linked semaphore to switch its state between ``locked'' and ``unlocked''. An unlocked semaphore lets one arrow pass and then switches its state to locked. A locked semaphore stops all arrows and can only unlock at the moment when a linked button is triggered. 

{\em Packages represent resources}. When an arrow passes over a package, it will automatically be picked up and delivered when passing over a delivery point.  To increase the complexity of CPP problems the game can express, {\em Parallel} contains three different types of packages to represent different types of shared resources in CPP. 

{\em Directional Switches and Diverters represent conditional statements.} Both elements direct arrows at intersections of tracks in different directions. Similar to a semaphore, a directional switch can link to a button. 
By contrast, the diverters direct arrows based on the type of packages they carry. Thus, directional switches represent conditional (if-then-else) statements depending on program variables (which can change via buttons), while directional switches represent conditions depending on the shared resources the threads are using (represented by the arrows carrying packages).

{\em Exchange Points represent Message Passing}. Exchange points are placed on the track and only appear in linked pairs. When an arrow arrives at an exchange point, it waits until another arrow arrives at the other linked exchange point. Then they swap packages if they are carrying any. This exchange is used to model problems that require message passing.

\subsection{Practicing CPP Problem Solving through Game Dynamics}

These basic formal elements and mechanics were designed specifically to give rise to game dynamics representing three fundamental CPP concepts: non-determinism, synchronization, and efficiency. 

{\em Non-determinism.}
Non-determinism is an essential characteristic of CPP and what makes CPP challenging even to skilled programmers familiar with sequential programming.
To capture this challenge, the key mechanics to drive non-determinism in {\em Parallel} are the random speeds of each arrow, which can vary at each simulation. The speed variation ensures solutions that work once may not succeed in the next run (as is true in actual parallel programming). The player can test her solution with the ``test'' function (bottom button in Fig.~\ref{fig:level06}), which will run her solution with a different configuration of the arrow speeds each time. When she feels ready, the player can ``submit'' her solution. 


{\em Synchronization.}
Concurrent and parallel programming involves addressing a set of synchronization challenges that do not arise in sequential programming such as {\em deadlocks}, {\em race conditions}, or preventing {\em starvation}. All of those concepts have their equivalent in {\em Parallel}. 
%
For example, the right-hand side diagram in Fig \ref{fig:level03} shows the critical section of this level: if two arrows get in there at the same time, correct execution cannot be ensured. So, the player needs to place semaphores and buttons to prevent that from happening. 

Similarly, starvation, deadlock, and other CPP concepts have their visual analogies in {\em Parallel}.

{\em Efficiency.}    
Like in CPP, the simplest way to deal with non-determi\-nism and synchronize threads is to block arrows in the game so that they move one at a time in deterministic ways. In other words, the problem becomes sequential. However, this approach is undesirable because it forgoes the benefits of CPP --- running multiple threads in parallel can boost efficiency. 
We implemented a star system into the game, which scores the solutions provided by students from one to three stars based on their efficiency. 

 \begin{table*}[tb]
    \caption{Topic specific self-efficacy survey questions (all questions were given on a Likert style scale of 1 - 10).}
    \label{tbl:survey}
    \centering
      \begin{tabular}{|l|l|l|}
      \hline
{\bf Question} & {\bf Meaning of a 1} & {\bf Meaning of a 10} \\ \hline
Q1. How well do you think you understand parallel programming? & I don't know anything about it & I know it inside and out \\ \hline
Q2. How much do you think you know about Critical Sections? & Nothing & Everything \\ \hline
Q3. How comfortable are you with spotting Critical Sections? & Cannot spot them & Could spot them easily \\ \hline
Q4. How much do you think you know about Starvation? & Nothing & Everything \\ \hline
Q5. How comfortable are you with spotting Starvation? & Cannot spot them & Could spot them easily \\ \hline
Q6. How much do you think you know about Race Conditions? & Nothing & Everything \\ \hline
Q7. How comfortable are you with spotting Race Conditions? & Cannot spot them & Can spot them easily \\ \hline
    \end{tabular}
\end{table*}

\subsection{A Sample Level}
Each level in {\em Parallel} encodes a carefully designed puzzle (e.g., Figure \ref{fig:level03}, Left), some directly based on the exercises in the classic CPP textbook {\em The little book of semaphores}~\cite{downey2008little}. The following example illustrates the type of CPP problems the game presents to the players. 

In Level 3 (Figure \ref{fig:level03}, Left), students will practice solving problems related to the {race condition}, a common type of challenge in CPP. The goal of the level is to design a synchronization mechanism, adding to the existing game elements already placed on the tracks, so that the two arrows can each take at least two packages from the package pickup point and deliver them to drop-off points. All tracks are directional and an arrow can only traverse in the direction specified (Figure \ref{fig:level03}). To reduce visual clutter, the game only displays directional information of all the tracks when the player turns on the function. 

The main challenge in this level is that a new package will not be regenerated at the pickup point until another package is dropped off. So if an arrow reaches the pickup point before the other arrow has dropped its package, the player will fail the level. This is because without a package, an arrow will be directed by the diverters to one of the two dead-end infinite loops at the top or bottom of the level. In other words, in order to solve this level, the player must solve the {\em race condition} between the two arrows. 

While difficult to solve by trial and error, players can use CPP concepts to complete this level. The easiest way is to first identify the {\em critical section} (highlighted in red in the right-hand side of Figure \ref{fig:level03}), and block it with semaphores and signals so that only one arrow can be inside of the critical section at a time. This solves the race condition, and thus the level. A possible solution that does exactly this is shown in Figure \ref{fig:level03}. The player needs to place two semaphores at both entrances to the critical section, and then two buttons at the two exits. She then needs to connect the buttons to the semaphores.

\section{Methods}\label{sec:methods}
In order to answer our research questions, we conducted a user study using concurrent mixed methods design~\cite{creswell2002educational}. The recruited students ($n=44$) participated in the study in a lab-style classroom, where each student had a computer with {\em Parallel} installed. They first completed a pre-study questionnaire.  Finally, all the students were asked to complete a post-questionnaires about their self-efficacy and user experience. Below are details of our methodology\footnote{ Anonymized dataset containing the student play through logs of all the levels in our study can be downloaded from [blinded for peer review].}. 


\subsection{Sample Selection}
We recruited 44 volunteers (41 male and 3 female) from two separate sections, Section A (n=25) and Section B (n=19), of the same Systems Programming course. It is a required course for sophomores or juniors in the Computer Science undergraduate program at a university at a major U.S. city in the Mid-Atlantic. The average age of participants is 21.11. 

Sections A and B followed exactly the same curriculum, and the classes met on different days of the week. At the time of our user study, students in Section B received a one-hour lecture on CPP basic concepts, such as running processes with multiple threads to increase efficiency, less than a week ago. By contrast, students in Section A had not yet been formally introduced to these concepts. 



\subsection{Pre-session Survey}
After giving their informed consent, all participants in the study were given an identical pre-survey to assess their perceived understandings of their abilities to solve CPP problems. These questions (Table~\ref{tbl:survey}) were administered as Likert items on a scale from 1 to 10 and asked students to rate their proficiency to solve specific tasks, such as identifying {\em race conditions} or {\em starvation} among other concepts, and their general capacity to solve CPP problems. These concepts were covered in the lecture which students in Section B had received several days ago. 
The survey was designed to reveal students' beliefs about their overall abilities and competence in solving concurrent and parallel programming problems (self-efficacy). 

\subsection{Gameplay Session}
All participants were given one hour to play five levels (Level 1, 2, 3, 4 \& 6) of {\em Parallel} in a fixed order on their own. Among them, a randomized subset ($n=6$, 5 male and 1 female, average age 22.83) 
was selected to take part in individual think-aloud sessions~\cite{ericsson68} to provide additional data on their problem-solving process. The rest of the students played the game without think-aloud. 

The selected levels had an increasing level of difficulty. Levels 1 and 2 primarily focused on introducing the core mechanics and the UI of the game. In the rest of the levels, the students need to solve {\em race conditions}, prevent {\em starvation} and identify and protect {\em critical sections} in order to {\em synchronize} the different arrows and accomplish the level goals.

Each of the participants in the think-aloud group played the game in a separate room with a researcher, who observed the student and prompted them for think-aloud. Screen and voice recordings of the think-aloud sessions were captured for gaining a detailed understanding of learners' problem-solving processes. Four of these students (3-3, 3-2, 3-3, 3-4) belonged to Section A, and two of them (5-3, 5-4) to Section B.


\subsection{Post-session Survey}
The post-session survey, identical to the pre-session survey, was administered to all participants at the end of the study. 

\subsection{Analysis}
Data analysis for this study began with qualitative coding of transcripts of students' think-aloud sessions~\cite{miles2014qualitative}. We used descriptive coding where members of our research team read over interview transcripts and made notes of any excerpts that stood out as being instances where students were showing self-efficacy, or where students gave an indication of the problem-solving strategy they were employing to solve the level. The next step entailed revisiting the transcripts and grouping these codes based on different themes. The team resolved differences between individual researchers' coding through discussions.

For our quantitative data, we used descriptive statistics and statistical significance tests. We used the paired $t$-test and the Mann-Whitney U tests to determine statistical significance as the samples we are comparing are paired (we compare pre- and post-test responses).


 \begin{table}[tb]
    \caption{Average (and standard deviation) reported self-efficacy from pre- to post-surveys ($n=44$), with $p$ values according to a paired $t$-test, and a Mann-Whitney U test.}
    \label{tbl:efficacy-pre-post}
    \centering
      \begin{tabular}{c|c|c|c|c}
      
 & {\bf Pre} & {\bf Post} & {\bf $t$-test} & {\bf M-W} \\ \hline
Q1  &   2.89 (1.54)     & 4.70 (1.64) & $7.4\times10^{-7}$ & $< 10^{-5}$ \\
Q2  &   2.25 (1.78)     & 3.66 (1.95) & $0.0001$ & $0.0004$ \\
Q3  &   1.95 (1.35)     & 3.55 (1.93) & $6.2\times10^{-6}$ & $< 10^{-5}$ \\
Q4  &   2.45 (1.99)     & 3.23 (1.98) & $0.01$ & $0.0524$ \\
Q5  &   2.00 (1.51)     & 3.16 (2.03) & $7.2\times10^{-5}$ &  $0.0056$ \\
Q6  &   2.98 (2.36)     & 3.91 (2.22) & $0.0003$ & $0.0232$ \\
Q7  &   2.61 (1.99)     & 3.80 (2.19) & $5.6\times10^{-6}$ & $0.0051$ \\ \hline
Avg. & 2.45 (1.84)   & 3.71 (2.04) & $<10^{-10}$ & $< 10^{-5}$ \\ \hline

Section A Avg. & 1.74 (1.52)   & 3.12 (2.06) & $<10^{-10}$ & $< 10^{-5}$ \\ 

Section B Avg. & 3.38 (1.80)   & 4.50 (1.72) & $<10^{-8}$ & $< 10^{-5}$ \\ \hline
    \end{tabular}
\end{table}

\section{Results}\label{sec:results}
Transcripts from student think-aloud gameplay and survey data revealed findings in relation to how self-efficacy impacted their gameplay and their thought progression in solving concurrent and parallel problems within the game, as well as how students' perceived self-efficacy changed after gameplay. Our results can be grouped into three main findings: (a) self-efficacy increased after playing the game; (b) the problem-solving strategies employed by students can be classified in three main types; and (c) self-efficacy is correlated with the percentage of time students spend in one of these problem-solving strategies (multithreaded problem solving). We elaborate on each one of these findings below.

 \begin{table*}[tb]
    \caption{Examples of talk coded as trial-and-error, single-threaded, and multithreaded problem solving}
    \label{tbl:talk}
    \centering
      \begin{tabular}{|p{0.3\textwidth}|p{0.3\textwidth}|p{0.3\textwidth}|}
      \hline
               {\bf Trial and Error} & {\bf Single-Threaded}   & {\bf Multithreaded}  \\ \hline

    {\em ``I am going to try something random, well because...''- Student 3-3}
    &   
    {\em ``I think the yellow one (student referring to just one thread) is probably gonna be simpler to figure out because, um, I know that it's just going in circles and I want to prevent that from happening'' - Student 5-4}    
    &   
    {\em ``So I guess the better thing to do would be to place the switch at each of these. So that way it will change to this on package delivery that would allow the next arrow to go through here and alternate it'' - Student 5-3}  \\ \hline
    {\em ``So let's see what happens.. Oh, I got lucky! Okay. Yeah.''- Student 3-1}
    &
    {\em ``I messed up the pink one (referring to just one thread). It should've gone back and delivered this again. So I need to change where it is''- Student 3-4}
    &
    {\em ``So I need him to give this block to someone else I guess-so he needs to give a normal block to that guy'' - Student 5-1}
    \\ \hline
    {\em ``I'm just gonna start test''- Student 3-2}
    &
    {\em ``I'll focus on this one first (referring to just one thread)'' - Student 5-1}
    &
    {\em ``Okay when it delivers the package and there will be a button here and that will switch this one (student discussing interactions between two arrows or threads)'' - Student 3-3}
    \\ \hline
    \end{tabular}
\end{table*}


\subsection{{\em Parallel} Increases Self-Efficacy}

The first thing that our quantitative analysis revealed is that students showed an increase in reported self-efficacy after playing the game. Table \ref{tbl:efficacy-pre-post} shows the average and standard deviation of the answers to the quantitative survey questions for all the students ($n = 44$) in our study.

As the table shows, the average response on the pre-test was 2.45, and it was 3.71 on the post-test. The difference was shown to be statistically significant according to both a paired $t$-test and a Mann-Whitney U test. In fact, the differences were found statistically significant at a $p<0.05$ level if we look at each individual question as well (except for question Q4, where we obtained a $p$ value of $0.0524$ according to the Mann-Whitney U test). We would like to emphasize that, as it can be seen in Table~\ref{tbl:survey}, the questions asked to the students concerned concurrent and parallel programming concepts rather than game-related concepts. The game does not include explicit descriptions of concurrent and parallel programming concepts, and thus, students need to make the connection between CPP concepts and game concepts themselves. For example, even if a certain level illustrates a {\em Race Condition}, this is never labeled as such in the game. Thus, the increased self-efficacy indicates that one hour of gameplay makes them feel more confident about CPP concepts than before playing. 

Another interesting result is that the $p$ values for questions Q2, Q4, and Q6 were higher than for the other questions. These are the questions of the form ``How much do you think you know about XXX?''. In questions that started with ``How comfortable are you with spotting XXX?'' the $p$ values were very low. This result matches our expectations, since the game provides the students with opportunities to gain practical skills (by practicing solving puzzles), but does not necessarily give them any additional theoretical knowledge of the concepts. This reinforces previous results by Zhu et al., showing that students were able to make the connection between the game concepts and CPP concepts in {\em Parallel}~\cite{zhu2019programming}.

Finally, the bottom two rows of Table~\ref{tbl:efficacy-pre-post} show the average results along with all questions when broken down by students from Section A (who received the CPP lecture after the study) and from Section B (who received the CPP lecture before the study). As expected, the self-efficacy answers for students from Section B are higher on average from those in Section A, as Section B students received a CPP lecture. Notice that the very low scores on the pre-survey for Section A are expected, as these students might have never even heard of terms such as {\em race condition}. 

Moreover, although playing the game increased the self-efficacy scores of both students in Section A and Section B, given that the game did not actually {\em explain} the concepts, but only made students work with them, among the qualitative answers provided by the students we noticed a significant difference in the answers of Section A and Section B. For example, even if they reported higher self-efficacy, a typical answer for a student in Section A was {\em ``The game didn't really explain the `concepts' just made you work through them.''} or {\em ``I think that once I am shown the traditional explanation it will all click for me.''}, whereas answers of the style of {\em ``Cleared things up a lot!''} were common in Section B. This indicates that, not unexpectedly, {\em Parallel} can be more useful after a minimal set of concepts have already been explained to the students.

\subsection{Problem Solving Strategies}

\begin{figure*}[tb]
    \centering
    \includegraphics[width=\textwidth]{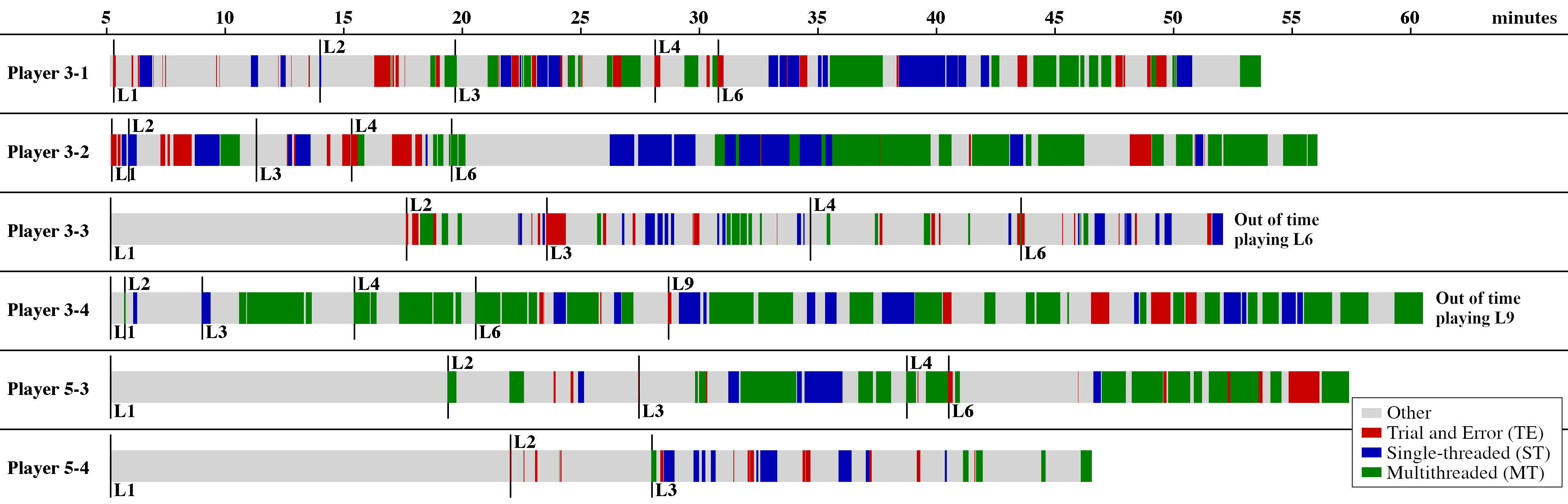}
    \caption{Time spent by students in trial and error/single-threaded/multithreaded problem solving by game level.}
    \label{fig:probsolving_by_level}
\end{figure*}

The think-aloud data collected with the students ($n=6$) offered an opportunity to analyze their reasoning process more closely. One theme that stood out was students deploying different {\em problem-solving strategies}. Using grounded theory, we identified three main strategies used by our participants: {\em trial-and-error}, {\em single-threaded problem solving}, and {\em multithreaded problem solving} (Table~\ref{tbl:talk}).

{\em Trial-and-error problem solving (TE)} was defined as any instance where students stated that they were not sure what they were doing but wanted to submit their answer to see if it worked. An example of this strategy is ``{\em So let's see what happens... Oh, I got lucky! Okay. Yeah.}'' (Student 3-1)  

{\em Single-threaded problem solving (ST)} was defined as any instance where students were talking about or focusing on one thread to solve the level or to solve the problem that they were currently facing. An example of this strategy is ``I'll focus on this one [thread] first.'' (Student 5-1)

{\em Multithreaded problem solving (MT)} was defined as any instance where students were talking about multiple threads or discussing that multiple threads were facing. An example is ``Okay when it delivers the package and there will be a button here and that will switch this one'.' (Student 3-3). In this case, the student was discussing the interactions between two arrows.

After the three problem-solving strategies were identified, the researchers used it to code the screen-recordings of the six participants' entire gameplay session. We triangulated the screen-recordings with the participants' think-aloud transcript to identify the sequence of gameplay actions associated with each problem-solving strategy. 

Figure \ref{fig:probsolving_by_level} shows the results of our analysis on the time each student spent on the three strategies. The horizontal axis represents the progression of time in minutes. We marked the beginning of each game level. All players were asked to play the 5 required levels (1, 2, 3, 4, and 6). Player 3-4 also attempted the much harder Level 9, after completing the required levels within half of the allocated one hour.  

When we determined that the player was using one of the three problem-solving strategies in his/her gameplay, we highlight that sequence of player actions in the respective color. However, the problem-solving strategy deployed wasn't always clear. The gray block represents that 1) the player did not provide enough talk-aloud information for the researchers to identify the problem-solving strategy, 2) the researchers cannot agree on which strategy was used, or) the student was doing something else other than solving the puzzle directly (e.g., reading the help menu). All players completed all the levels they started except Player 3-3, who ran out of time playing level 6, and Player 3-4, who ran out of time playing level 9.

\begin{figure}[tb]
    \centering
    \includegraphics[width=\columnwidth]{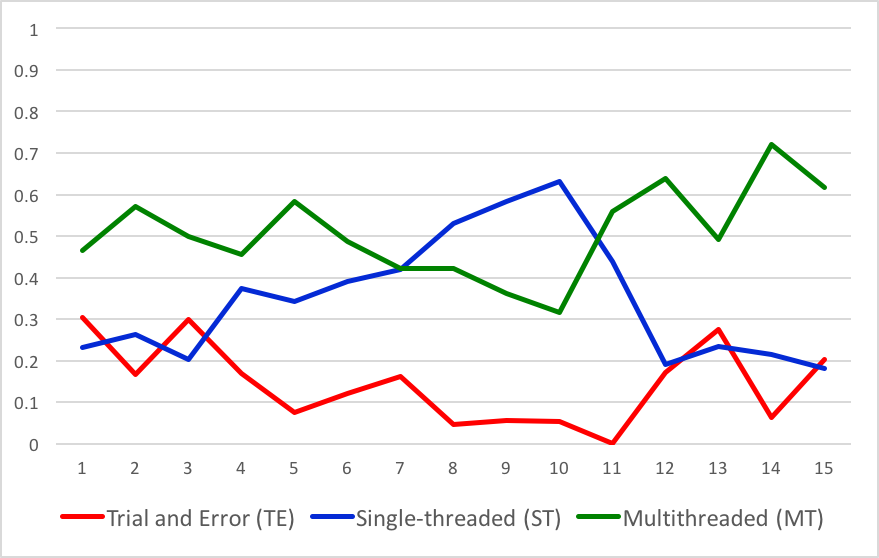}
    \caption{Proportion of time students spent deploying each of the three strategies over time in each level. Horizontal axis represents minutes since the start of the current level.}
    \label{fig:probsolving_by_time}
\end{figure}

Figure \ref{fig:probsolving_by_time} shows an aggregated view of the proportion of time students spent in each type of problem-solving strategy. The horizontal axis shows minutes since the start of a level, and the vertical axis shows the proportion of each problem-solving strategy according to our coded data. This was generated by aggregating the labels for all 6 students over all the levels they played. We only show the first 15 minutes, since few levels require more than 15 minutes of time, and after this point, data is too sparse to be useful for analysis. The figure shows that trial and error is more common at the start of a level and declines slowly after that (with a second peak after 12 - 13 minutes, probably indicating that some students might be frustrated with the current level after a while, and revert to this strategy). Single-threaded problem solving is mostly deployed after 3 - 4 minutes and then decays after 10 minutes. Multithreaded problem solving slowly increases over time, indicating that as players spend more time on a level, they start leaning more and more towards this type of problem-solving. 

It is worth noting that the player that transitioned to multithreaded problem solving earliest (Player 3-4) was the only one to reach level 9 (L9). The calculated averages of time that students spent at each level of problem-solving, as seen in Figure \ref{fig:probsolving_by_level} are represented quantitatively in Table~\ref{tbl:scores}, showing that out of the three strategies, students spent most of their time in MT. 

\begin{table}[tb]
    \caption{Average percentage of time students spent in the TE/ST/MT problem solving stages. Sample includes students chosen for think-aloud sessions (n=6).}
    \label{tbl:scores}
    \centering
    \begin{small}
    \begin{tabular}{c|c|c}
        {\bf Trial and Error (TE)}    & {\bf Single-threaded (ST)}   & {\bf Multithreaded (MT)}  \\ \hline
    18.87\%  &   34.55\%    &   46.58\%    \\ \hline
    \end{tabular}
    \end{small}
\end{table}

\subsection{Connecting Self-Efficacy and Problem Solving Strategy}

\begin{table}[b]
    \caption{Pearson correlation coefficient between self-efficacy and time spend in each of the three problem solving strategies. Sample includes students chosen for think-aloud sessions (n=6).}
    \label{tbl:correlation}
    \centering
    \begin{small}
    \begin{tabular}{c|c|c|c}
        & {\bf TE}    & {\bf ST}   & {\bf MT}  \\ \hline
        {\bf Pre Self-Efficacy} & 0.26 & 0.04 & -0.13   \\ \hline
        {\bf Post Self-Efficacy} & -0.34 & -0.57 & 0.56   \\ \hline
        {\bf Self-Efficacy change} & -0.58 & -0.44 & 0.56 \\ \hline
    \end{tabular}
    \end{small}
\end{table}

After we calculated the percentage of time a student spent at each of the three problem-solving strategies, 
We compared this information with quantitative data gathered from student pre- and post- surveys of self-efficacy.
%
%
Based on the quantitative data gathered from the survey and the qualitative TE/ST/MT theme coding within student think-alouds, it was evident that there was a connection between high student self-efficacy on the pre-survey and their time spent in multi-threaded problem-solving. 

Table~\ref{tbl:efficacy-progression} shows each of the six student's average scores from the pre- and post- surveys in connection to the percentage of time she/he spent on the three problem-solving strategies,  and the number of levels they managed to complete during the study. The table shows that the two students with less fraction of the time spend at MT were the only ones not able to solve all 5 levels in the study. The table also shows that students who showed higher self-efficacy than their peers after gameplay were more likely to spend time at the multi-threaded (MT) level of problem-solving, which for the game {\em Parallel} was the ideal stage in the progression. 

This is more clearly seen in Table \ref{tbl:correlation}, which shows the Pearson correlation coefficient between the pre- and post-survey self-efficacy and the percentage of time each student spent in each of the three problem-solving strategies. As the table shows, post-survey self-efficacy and also the change in self-efficacy between pre- to post- surveys, show a strong correlation with the problem-solving strategies. For example, the correlation between post self-efficacy and MT is 0.56, which denotes a rather strong correlation (students with high self-efficacy tend to spend more time deploying the MT strategy). A very strong negative correlation also exists between post self-efficacy and ST. The same strong correlation is present when we consider the change in self-efficacy between pre- and post-surveys. This makes sense and indicates that students with higher self-efficacy spent less time in single-threaded problem solving and more on multithreaded problem-solving. While we acknowledge that this analysis is based on a small sample of six students, we believe it shows a promising trend that deserves further investigation with a larger sample.

\begin{table}[tb]
\centering
\caption{Connection between self-efficacy and  TE/ST/MT stages of problem solving. Sample includes students chosen for think-aloud sessions (n=6)}
\label{tbl:efficacy-progression}
\begin{small}
\begin{tabular}{c|c|c|c|c|c|c}

& \multicolumn{2}{c|}{\textbf{Self-Efficacy}} & \multicolumn{3}{c|}{\textbf{Problem Solving (\%)}}  & \multicolumn{1}{c}{}\\ \hline
\multicolumn{1}{l|}{ID} & Pre                  & Post                 & TE              & SE              & MT        & Levels Solved     \\ \hline \hline
\multicolumn{1}{l|}{3-1}        & 1.29                 & 2.14                 & 18.68\%           & 41.95\%           & 39.37\%     & 5     \\ \hline
\multicolumn{1}{l|}{3-2}        & 1.43                 & 2.71                 & 14.57\%           & 44.94\%           & 40.49\%     & 5      \\ \hline
\multicolumn{1}{l|}{3-3}        & 2.00                 & 2.14                 & 31.61\%           & 36.44\%           & 31.95\%     & 4      \\ \hline
\multicolumn{1}{l|}{3-4}        & 1.14                 & 3.00                 & 9.25\%            & 19.87\%           & 70.88\%     & 5      \\ \hline
\multicolumn{1}{l|}{5-3}        & 4.57                 & 4.71                 & 15.63\%           & 14.27\%           & 70.10\%     & 5      \\ \hline
\multicolumn{1}{l|}{5-4}        & 5.71                 & 3.43                 & 23.49\%           & 49.84\%           & 26.67\%     & 3      \\ \hline
\end{tabular}
\end{small}
\end{table}

Finally, Table \ref{tbl:sample-validation} shows the average pre- and post- reported self-efficacy for students that participated in the think-aloud session ($n=6$) and those who did not ($n=38$), to show that the think-aloud group exhibited similar statistical trends to the larger group of participants. We note that post self-efficacy seems lower in the think-aloud group than in the general group. However, notice that the think-aloud group included 4 participants from Section A and only 2 from Section B, and thus the average is heavily biased towards Section A self-efficacy values, which were lower.

\begin{table}[tb]
    \caption{Average (and standard deviation) reported self-efficacy scores of think-aloud participants and regular participants, and average number of levels solved.}
    \label{tbl:sample-validation}
    \centering
    \begin{small}
    \begin{tabular}{c|c|c|c|c}
        & {\bf Pre}    & {\bf Post}   & {\bf Levels Solved} & {\bf n}  \\ \hline
        {\bf Think-aloud} & 2.69 (2.02) & 3.02 (1.42) & 4.50 & 6   \\ \hline
        {\bf Not think-aloud} & 2.41 (1.81) & 3.82 (2.10) & 4.82 &  38   \\ \hline
        {\bf All} & 2.45 (1.84) & 3.71 (2.04) & 4.75 & 44 \\ \hline
    \end{tabular}
    \end{small}
\end{table}

\section{Discussion}\label{sec:discussion}
In this section, we interpret our results to answer the two general questions. 

\subsection{Self-Efficacy}
General question 1: {\em How does {\em Parallel} impact the learners' level of self-efficacy, if at all?}

Research in the past has shown that self-efficacy has a connection to student learning and motivation~\cite{bandura1977self,collins1982self,schunk2012motivation}. Our study reinforces such a result by showing that students' gameplay was influenced by their prior self-efficacy, and conversely that gameplay affected their self-efficacy regarding CPP concepts such as critical sections, starvation, and race conditions. It also showed that the higher a student's self-efficacy was prior to their playing the game, the more time they would spend in multithreaded problem solving, a strategy that for this game shows us that students are thinking in a parallel or concurrent fashion within the game. This study also reinforces previous results showing that games can potentially increase self-efficacy~\cite{ketelhut2007impact,meluso2012enhancing,wang2010effects,jui2011game}, and specifically shows that our design of {\em Parallel} was able to do so in the context of concurrent and parallel programming.

Despite the small sample size for the think-aloud protocol, this study provides several implications for practice as well as next steps for the future work around {\em Parallel} as well as self-efficacy. Through our results, we found that students are, in fact, improving in self-efficacy after having played the game {\em Parallel} and that self-efficacy is a good predictor of students' abilities to consider multiple threads when playing the game rather than relying solely on trial-and-error. This suggests that this game has the potential to help to motivate and maintain students' interest in concurrent and parallel programming. This may aid in boosting student self-esteem and persistence around challenging programming concepts. 

\subsection{Problem-Solving Strategies}
General question 2:{\em What kind of problem-solving strategies do the learners use in the context of playing {\em Parallel}?}

Through the grounded theory methodology, we identified three main problem-solving strategies:  trial-and-error (TE), single-threaded (ST), to multi-threaded (MT) problem-solving. Using this framework, our results suggest that most students would progress from trial-and-error to multi-threaded problem solving within the game, reverting back to Trial-and-error or single-threaded problem solving when they were having difficulty with the higher difficulty levels (usually expressing this out loud). This finding is consistent with literature \cite{lonnberg2009students}. As Lonnberg argued, ``Some trial and error is inevitable, as one can not understand the development task fully until one has attempted it, making it hard to design ahead.'' In the case of {\em Parallel} we observed some successful students run the level before placing any game elements, in order to understand the problem and develop problem-solving strategies.

Based on student feedback from the qualitative think-aloud, we also know that students often moved away from trial-and-error as they became more familiar with the games user interface (UI), and moved up towards multi-threaded problem solving within the game as they began to think in a more concurrent fashion. This progression can be seen in Figure \ref{fig:probsolving_by_level}, which shows the quantified time that each student spent at each level within the progression in comparison with each other. 

The game level progression is designed to increases difficulty regarding the CPP problems they embody. This means that it will be increasingly difficult to complete the level with the trial-and-error strategy.  This is confirmed in our results. 
Initially, students begin to focus more on the actions of individual arrows in the game, essentially trying to manipulate single threads to achieve the goals of a level. As the levels get more complex, students need to consider relationships between multiple arrows and how to synchronize their behaviors. These three modes of problem-solving can be thought of as a progression where students first try random exploring the puzzles, move to independent observations of a single thread, and ultimately need to consider the range of possible interactions between multiple threads. Our belief is that increased time spent in the final, multithreaded stage is required for dealing with specific problems posed in the game and in concurrent and parallel programming more generally, but requires mastering several of the key underlying concepts.

What is particularly interesting is the connection between self-efficacy and problem-solving strategy. Since the problem-solving strategy is something that can be directly observed during gameplay, it is possible to automatically detect these strategy and thus infer self-efficacy. 


\section{Limitation}
The main limitation of our study was the sample size. Although survey data contains a larger set of students, think-aloud was only possible with a smaller set of students. Another limitation is due to the use of the think-aloud protocol. It is possible that having to verbally articulate their thinking progress may alter how these students played the game and their problem-solving process. However, since relatively little is known about how students learn CPP-related problem solving and systematic thinking skills, the think-aloud protocol offered a useful way to advance our knowledge. 
Finally, a potential limitation is the use of a purely visual game where the students do not code in a traditional way. It is conceivable that students may adopt a different learning process in games like {\em Parallel}. The generalizability of our results need to be further researched. 


\section{Conclusions and Future Work}

This paper has presented {\em Parallel}, an educational programming game designed to teach concurrent and parallel programming (CPP) concepts. Specifically, we reported on the results of a user study focusing on how self-efficacy impacts undergraduate students learning CPP concepts through {\em Parallel}. Our results identified: (a) self-efficacy increased significantly from the pre-survey to the post-survey; (b) three different problem-solving strategies (trial and error, single-threaded problem solving and multithreaded problem solving) and how students switch from one to another; and finally, (c) that higher self-efficacy before playing the game is correlated to a higher amount of time in multithreaded problem-solving.

As part of or future work, we plan to conduct further studies, aiming at increasing the sample size of our think-aloud protocols to better understand possible correlations between self-efficacy and problem-solving strategies. Specifically, we are interested in investigating whether the three problem-solving strategies identified in this study are general enough to characterize those of a larger student population. Our future studies will also examine how the increased self-efficacy that seems to come through playing the game might transfer to actual concurrent and parallel programming tasks.

\section{Acknowledgements}
We would like to thank Bill Mongan for collaborating with this study by giving us access to students in one of his classes. This project is partially supported by Cyberlearning NSF grant 1523116.

\bibliographystyle{ACM-Reference-Format}

\end{document}